\documentclass[aps,reprint,twocolumn,amsmath,amssymb]{revtex4-1}
\usepackage{graphicx}

\makeatletter
\def\ps@pprintTitle{%
 \let\@oddhead\@empty
 \let\@evenhead\@empty
 \def\@oddfoot{}%
 \let\@evenfoot\@oddfoot}
\makeatother

\begin{document}



\title{Impact of Galactic magnetic field modelling on searches of point sources via UHECR-neutrino correlations}

\author{J.A. Carpio and A.M. Gago}
\address{Secci\'on F\'isica, Departamento de Ciencias, Pontificia Universidad Cat\'olica del Per\'u, Apartado 1761, Lima, Per\'u}


\begin{abstract}
We apply the Jansson-Farrar JF12 magnetic field configuration in the context of point source searches by correlating the Telescope Array
ultra-high energy cosmic ray data and the IceCube-40 neutrino candidates, as well as other magnetic field hypotheses. Our field hypotheses 
are: no magnetic field, the JF12 field considering only the regular component, the JF12 full magnetic field, which is a combination of 
regular and random field components, and the standard turbulent magnetic field used in previous correlation analyses. As expected from a 
neutrino sample such as IceCube-40, consistent with atmospheric neutrinos, we have found no significant correlation signal in all the cases. 
Therefore, this paper is mainly devoted to the comparison of the effect of the different magnetic field hypotheses on the minimum neutrino 
source flux strength required for a $5\sigma$ discovery and the derived $90\%$ CL upper limits. We also incorporate in our comparison the 
cases of different power law indices $\alpha=2,\alpha=2.3$ for the neutrino point source flux. Our results indicate that the discovery 
potential for a point source search is sensitive to changes in the magnetic field assumptions, being the 
difference between the models mentioned before between 15\% and 26\%. Finally, as a collateral result, we implement
a novel parameterisation of the JF12 random component. \end{abstract}
\maketitle

\section{INTRODUCTION}
One of the most important quests in particle astrophysics is to identify the sources that produce ultra-high energy cosmic rays (UHECR). One 
way to achieve this goal is to directionally correlate UHECR with astrophysical neutrinos. This correlation would not only indicate us 
the sites of hadronic acceleration, but also their non single-shot transient nature. On the other hand, a positive correlation signal 
requires sources producing a similar order of fluxes of UHECR (e.g.protons) and neutrinos. This condition is fulfilled by sources with
a proton interaction opacity $\tau \gtrsim 1$. Meanwhile, sources with $\tau \ll 1$ or $\tau \gg 1$ would either produce almost exclusively protons with a small associated neutrino flux or produce only neutrinos, absorbing the corresponding protons, respectively. 
A mean free path length can be linked to the interaction between the protons escaping from the source and the 
Cosmic Microwave Background (CMB), defining the so called GZK sphere of radius $\sim 100$ Mpc ~\cite{Greisen66,Zatsepin66}. This would give us the limit of the farthest distance of observable UHECR sources, an implicit condition for any correlation analysis.

One key ingredient in a directional correlation analysis between UHECR and neutrinos is to have an accurate determination of the 
UHECR magnetic deflection. Since the UHECR are charged particles, in their travel to Earth, they are going to be deflected 
due to its interaction with the magnetic field inside and outside the Galaxy, the latter being true if they are coming from an extragalactic 
source.  

This type of analysis has been perfomed under different magnetic field hypotheses. Examples include correlation studies between UHECR from 
the Pierre Auger Observatory (PAO) and neutrino events from the ANTARES Telescope \cite{Martinez12} and between UHECR from Telescope Array 
(TA) and neutrino events from IceCube \cite{Fang14}. In all these analyses, an energy independent magnetic deflection has been used. 
This kind of estimation is reasonable since we are not certain about the description of the Galactic or the Extragalactic magnetic field. 
However, it is interesting to examine the impact on correlation analyses, calculating the magnetic deflections under different Galactic 
Magnetic field models, taking into account energy dependent considerations in their estimations. Following this idea, a study of cosmic ray 
deflections from Centaurus A was proposed in \cite{Keivani15} in which the more recent version of the Jansson and Farrar 
\cite{Jansson12a,Jansson12b} (JF12) magnetic field model has been implemented. The JF12 is an improved model of the Galactic magnetic field,
including a regular and random field component. This model fits very well the WMAP7 Galactic synchrotron emission map and more than 40,000 
extragalactic Faraday rotation measurements. The JF12 model also includes an out-of-plane component and striated-random fields.   

In this work we use four different magnetic field hypotheses: no field, the regular component of the JF12, the full JF12 field model, and the 
standard turbulent magnetic field. These hypotheses are initially used to study the correlation between the Telescope Array ultra-high energy 
cosmic ray data \cite{Abbasi14}  and the IceCube-40 neutrino candidates \cite{Abbasi11}. Since no correlation is expected, the main analysis 
of this paper is the comparison of the neutrino flux requirement for a $5\sigma$ discovery and the $90\%$ CL upper limits, under the four 
magnetic field hypotheses. The paper is divided as follows: in section 2 we describe the magnetic field models used in this study and we 
also present the parametrization that we have developed of the angular deflections from the random component of the JF12 field. 
In sections 3 and 4 we outline the method used to find correlations between UHECR and neutrinos and we describe how to include magnetic 
field deflections in our statistical analysis. In section 5 we present our results and in section 6 our conclusions. 

\section{GALACTIC MAGNETIC FIELD MODELS}
The angular deflection $\delta$ of an ultrarelativistic particle of charge $Ze$ and energy $E$ due to the magnetic field is described by:
\begin{equation}
\delta\propto \frac{Ze}{E}\left|\int \hat{\mathbf{p}}\times\mathbf{B}ds\right|
\end{equation}
This formula reflects the inverse proportionality between $\delta$ and $E$ which is a key issue for understanding the parameterisation
presented in section 3.

\begin{figure}[h]
\includegraphics[width=0.5\textwidth]{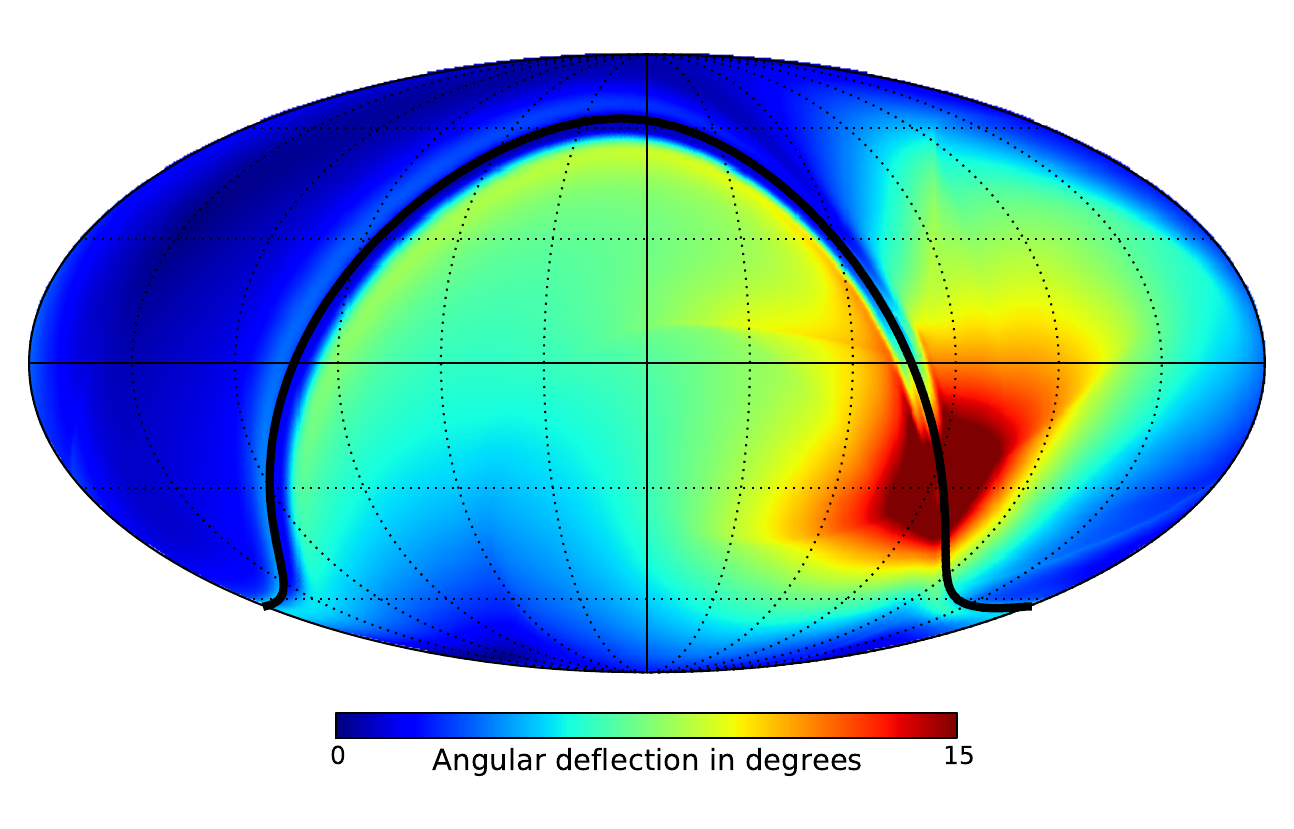}
\hspace*{-26pt}
\includegraphics[width=0.59\textwidth]{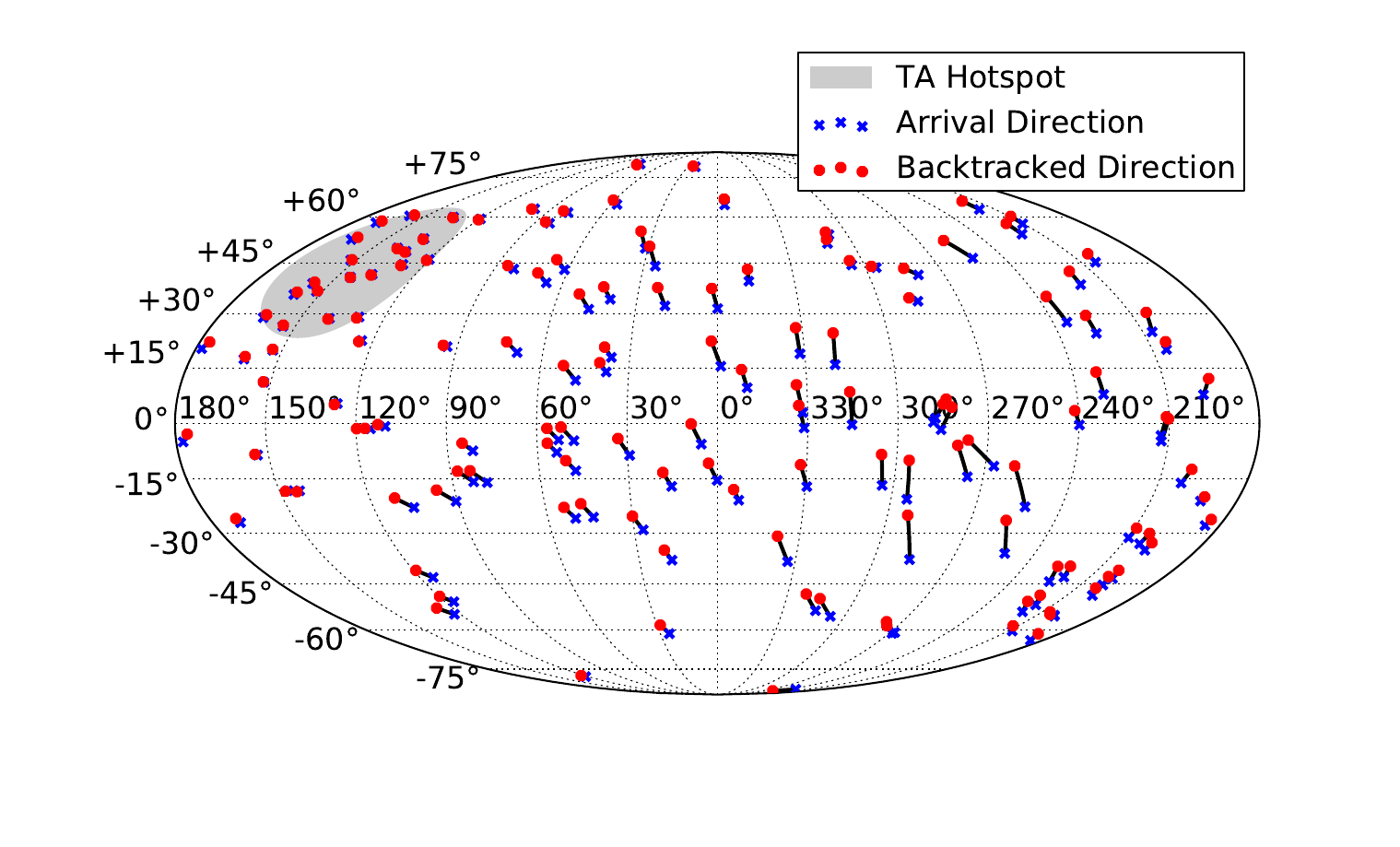}
\vspace{-30pt}
\caption{Top panel: Angular deflection from the regular component of the JF12 Field for a backtracked proton of energy $E=57$ EeV. The 
galactic plane is indicated by the thick black line. Map drawn using HEALPix \cite{Gorski05}. Bottom panel: Deflection of PAO and 
TA UHECR events under the regular component of the JF12 field. The shaded region marks the TA hotspot. Maps in equatorial coordinates.}
\label{Fig1}
\end{figure}

The deflection of a charged particle as it travels through a turbulent extragalactic magnetic field (EGMF), with strength of order $~6$nG, is 
proportional to $\sqrt{D}|Z|/E$ where $D$ is the source distance, $Z$ the charge number of the UHECR and $E$ its energy 
\cite{Dolag04,Kotera08}. These deflections are typically $\leq 1^\circ$ for protons with energy above 40 EeV and a propagation distance of 
order $\sim 1$ Gpc. Meanwhile, deflections caused by the galactic magnetic field (GMF), 
with strength of order $~6\mu$G, may be significant. In fact, this is particularly significant
if the cosmic rays are composed by heavy nuclei such as iron and may be deflected by a few tens of degrees even for 
energies above $10^{20}$ eV.

For distances up to 500 Mpc and energies $E\geq 4\times 10^{19}$ eV, typical deflections from EGMF are smaller than the resolution of UHECR 
detectors (see, e.g., \cite{Dolag04}). Thus, if the UHECR sources are within the GZK sphere of radius $\sim 100$ Mpc, we 
may neglect EGMF deflection for energies $E\geq 3\times 10^{19}$ eV, which is the case in our analysis.  

Our correlation analysis is mainly tested under two different Galactic magnetic field models. 
One of these models is the JF12 GMF model \cite{Jansson12a,Jansson12b}, designed to fit the rotation measures and polarized 
synchrotron data, as we have already mentioned before. The JF12 model separates the magnetic field into three components: regular (coherent), striated and purely random fields (random components). 
All components extend up to $r=20$ kpc, where $r$ is the in-plane radius with origin at the galactic center. The regular component (or 
large-scale field) is a superposition of three fields: spiral disk field, toroidal halo field and the X-shaped poloidal halo field. Figure
\ref{Fig1} shows the deflection from the  coherent field for a backtracked 57 EeV proton. Deflections are stronger for backtracked 
directions below the Galactic plane and are strongest when the particle propagates across the galactic center. 
  
In addition to the regular component of the JF12, the striated random field is fully aligned to all components of the regular field. The final
component is an isotropic random small-scale field. The coherence length of these fields is expected to be of order 100 pc or less. The GMF 
will be modelled using the best-fit parameters in \cite{Jansson12a,Jansson12b} and assuming a coherence length of 60 pc for the fully random 
component. 

The other model, which we denote by ``standard'', was used in Ref. \cite{Martinez12,Lauer11} for similar UHECR-neutrino correlation studies. 
In this model the deflections follow a Gaussian distribution with width $3^\circ$. In \cite{Aab14}, this method was changed to a Fisher 
distribution with an energy dependent concentration parameter. 

\subsection{Random component parameterization of the JF12 GMF}
\begin{figure*}[t!]
\begin{center}
\centerline{%
\includegraphics[width=0.5\textwidth]{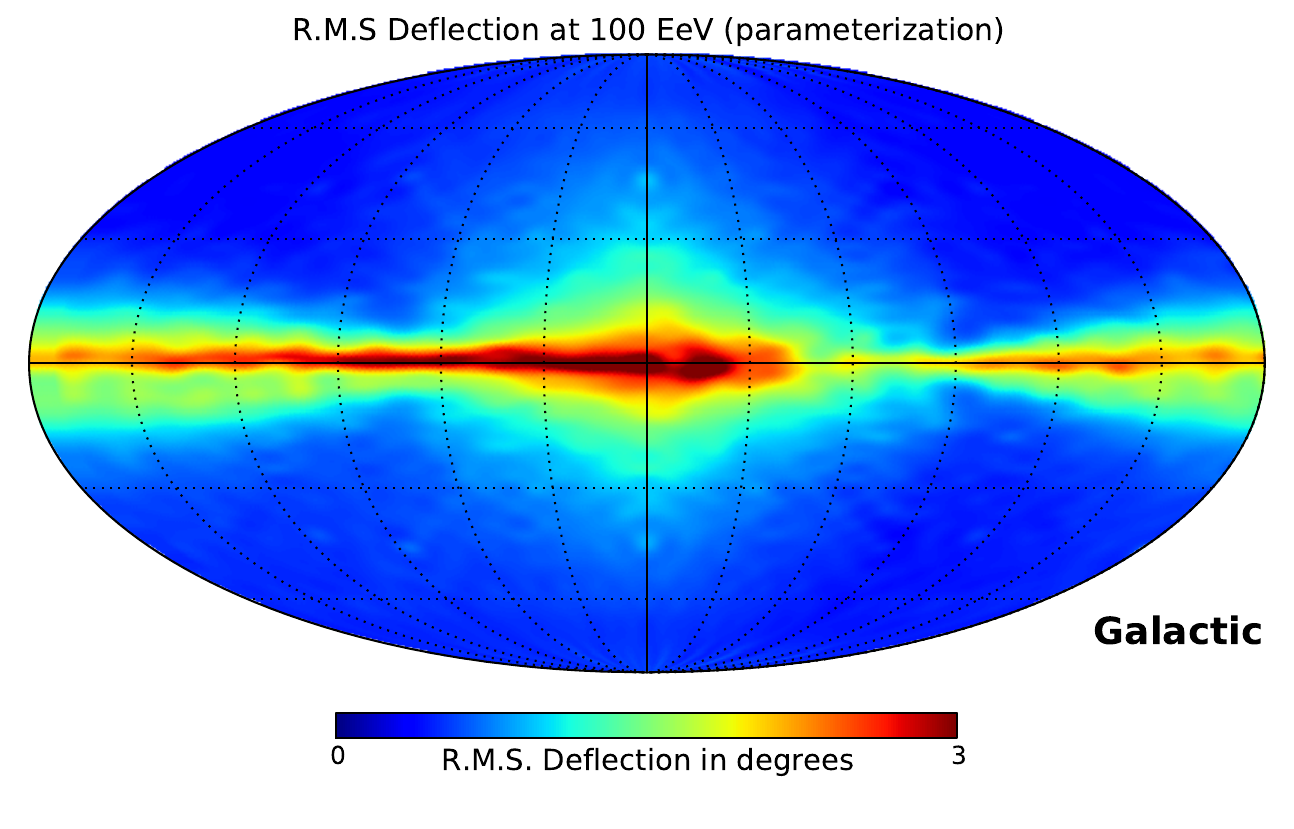}%
\includegraphics[width=0.5\textwidth]{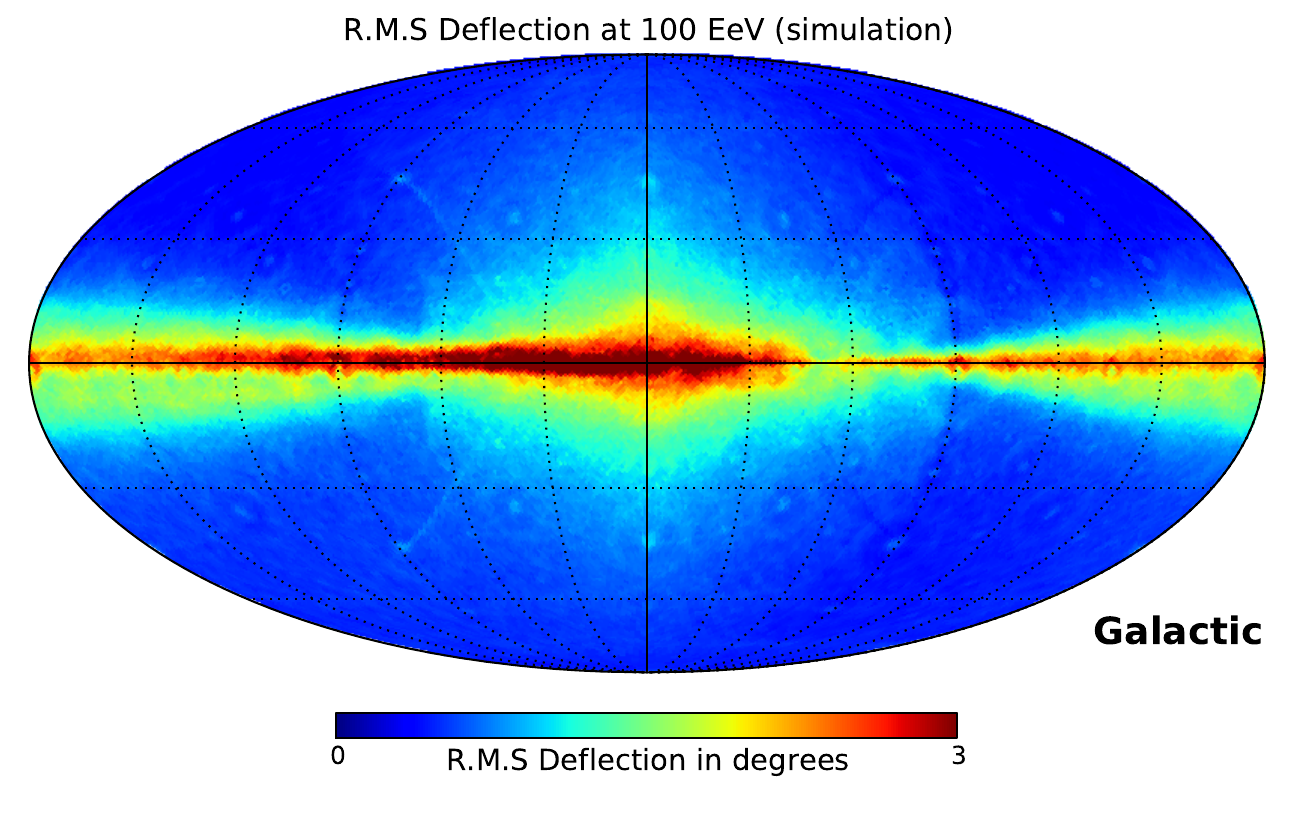}%
}%
\caption{Left panel: R.M.S. deflection due to random field components for 100 EeV protons, using our parameterization (Eq. (5)).
Right panel: The same plot by backtracking for several realizations of the full JF12 Field (simulation).
Map drawn using HEALPix \cite{Gorski05}.}
\vspace{-20pt}
\label{Fig2}
\end{center}
\end{figure*}

A secondary contribution of this work is the implementation of a parameterization of the random component of the JF12 GMF model. This 
analytic expression would help us in having a better insight into the behaviour of this component while saving computational time 
in the calculations of our analysis. This parametrization is valid for small deflections, a condition that restricts the energy range of 
UHECR, which are protons, to energies greater than 57 EeV.
We have chosen protons with this energy since the deflections are going to be small, which is a requirement for our results to be valid. In 
addition, for the backtracking method, we have used CRPropa 3 \cite{Batista13} to propagate cosmic rays under the JF12 magnetic field model, 
using the best fit values given in \cite{Jansson12a}, for the coherent field parameters, and in \cite{Jansson12b}, for the random field 
parameters.

\begin{figure}
\hspace*{-25pt}
\includegraphics[width=0.59\textwidth]{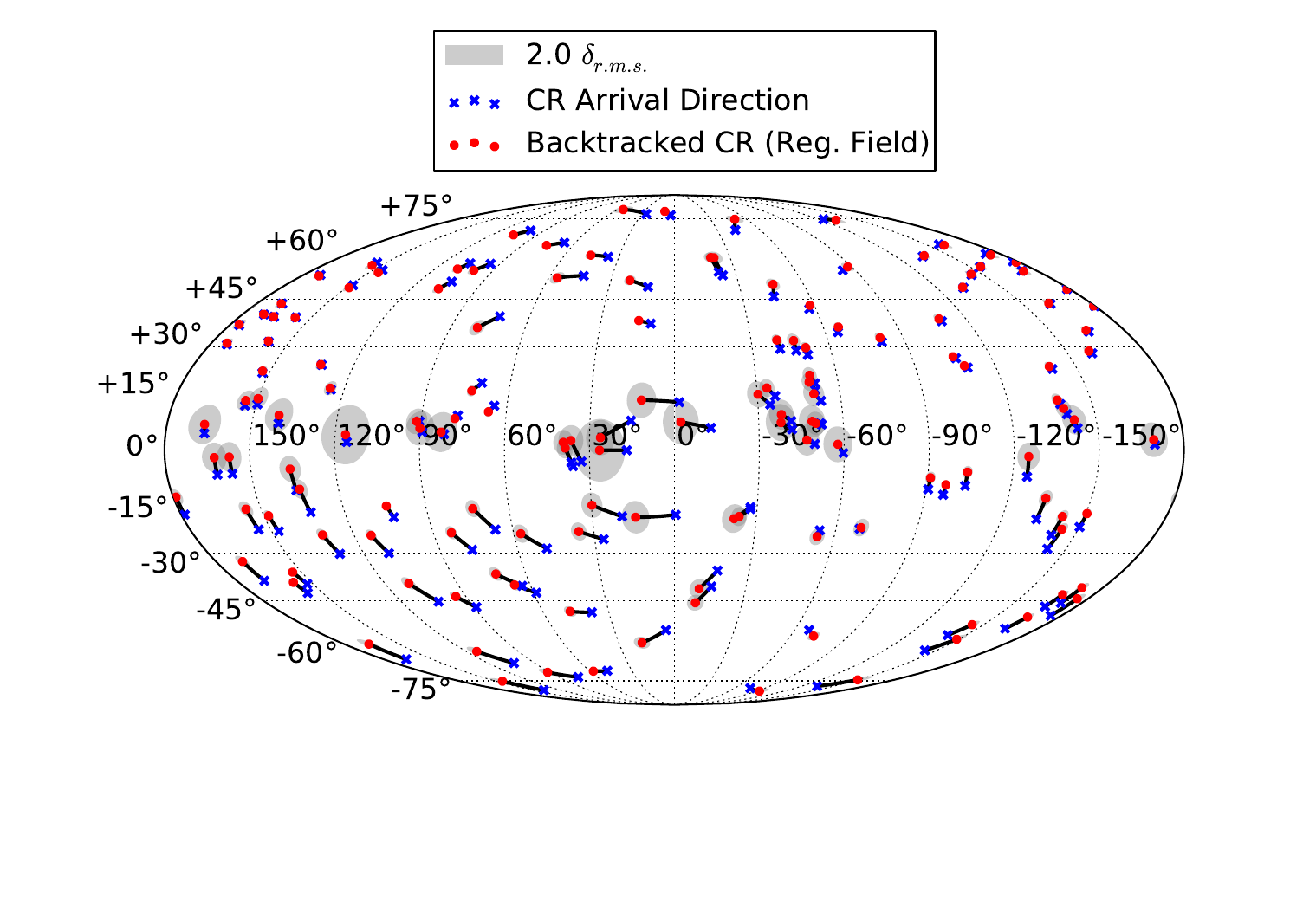}
\vspace{-50pt}
\caption{Deflection of the 69 UHECR reported by PAO in \cite{Abreu10} and the 72 TA events in Galactic coordinates. The blue crosses are the 
arrival directions of
the cosmic rays as seen from Earth. The red dots correspond to the backtracked events using the regular component of the JF12 Field and each
event is connected to its observed arrival direction by a black line. The gray circles have radii $2.0\;\delta_\textrm{r.m.s}$ (see Eq. (5)), 
which correspond to the additional deflection from the random components of the JF12 Field.}
\label{Fig3}
\end{figure}
Our procedure for obtaining the parameterisation goes as follows: firstly, for a 
given initial arrival direction (observed at the Earth) $P_0=(l_0,b_0)$ of a proton 
with energy $E$, where $(l,b)$ are in the galactic coordinate system with $-180^\circ\leq l<180^\circ$ the galactic longitude and 
$-90^\circ\leq b\leq 90^\circ$ the galactic latitude, we backtrack it through the coherent field to the position $P_c=(l_c,b_c)$. The latter 
is the final position reached by the proton before it leaves the GMF. Secondly, starting again from $P_0$, 
we backtrack another proton with 
the same energy $E$ through the full magnetic field (coherent plus random) until it arrives to its final position $P=(l,b)$, before it exits the GMF. 
Then the angular distance $\delta$ between $P_c$ and $P$ is calculated. After repeating this process over different realizations of the random 
component of the field, we may construct a distribution of the angular distance $\delta$.

To functionally describe the deflection, we have followed \cite{Aab14} where the random field deflections 
are given by a Fisher distribution. Under the small deflection angle hypothesis, the probability distribution function $f$ of the random field 
deflections can be described by a Rayleigh distribution:
\begin{equation}
f (\delta) = 2\lambda \delta \exp\left(- \lambda\delta^2\right)
\label{Parameterization}
\end{equation}
where $\lambda = \lambda(E)|_{l_0,b_0}$ is a fit parameter for a given $l_0,b_0$. We find that the optimal energy dependence of $\lambda$ is
of the form
\begin{equation}
\lambda|_{l_0,b_0}(E) = A_1|_{l_0,b_0} E+ A_2|_{l_0,b_0} E^2
\end{equation}
This energy dependence was tested for energies 57 EeV$\leq E\leq$200 EeV by taking 15 energies logarithmically spaced
in this interval. The parameters $A_1,A_2$ are given on a $50\times 50$ grid distributed uniformly across $-180^\circ\leq l_0<180^\circ$ 
and $-0.9999\leq\sin(b_0)\leq 0.9999$. Using the 2500 tabulated points as a sample, the weighted averages (weight $w_i=\cos(b_i)$) are given by
\begin{displaymath}
\langle A_1\rangle \sim (0.1\pm 3.0)\times 10^{-3}\textrm{EeV}^{-1} 
\end{displaymath}\begin{equation}
\langle A_2\rangle \sim (2.6\pm 1.9)\times 10^{-4}\textrm{EeV}^{-2}
\end{equation}
The distribution $f$ also satisfies
\begin{equation}
\langle \delta\rangle = \frac{\sqrt{\pi}}{2\sqrt{\lambda}}\qquad\qquad \delta_\textrm{r.m.s.} = \sqrt{\langle\delta^2\rangle} = 
\frac{1}{\sqrt{\lambda}}
\label{rmsdeflection}
\end{equation}
From Eq. (5) we see that the $\delta_\textrm{r.m.s}$ deflection decreases for higher energies being that 
for energies above 100 EeV it goes like $\lambda\approx A_2 E^2$.
The behaviour $\delta_\textrm{r.m.s}\propto E^{-1}$ is therefore recovered in the higher energy regime, in agreement with Eq. (1). In fact, 
this is the reason why there are no terms of order higher than $E^2$ in the expression for $\lambda$.
In order to test our proposed parameterization, we display in Figure \ref{Fig2} a comparison between this one and the numerical simulations,
which are based on Runge-Kutta methods for the $\delta_\textrm{r.m.s.}$ of 100 EeV protons. These simulations are performed over 640
realizations of the full JF12 field and with the sky divided into only 49152 arrival directions. It is clear the excellent agreement between
our parameterization and the numerical calculations.

Using $f$ we can reduce the computational time by up to two orders of magnitude, in comparison 
with the numerical calculations. The function $f$ only gives information on the angular deflection 
$\delta$, leaving for each $\delta$ an allowed region, an annulus, for the final backtracked position.
We assume a symmetrical distribution of the particles in the annulus.

In Figure \ref{Fig3} we show the effect of Galactic magnetic deflections on the 69 PAO UHECR events 
and 72 TA events. These deflections have been obtained due to the regular 
component, that produce a fixed separation, and the random component, depicted using our parameterization. 
It is evident how the greatest deflection given by the random component occurs mainly in the vicinity of 
the galactic plane.

\section{GENERAL SCHEME OF THE CORRELATION ANALYSIS}
In order to calculate the correlations between IceCube neutrinos and TA UHECRs, we have followed the source stacking method outlined in 
\cite{Martinez12}. This method adds up the flux intensities from a group of single sources concentrated in a small region of the sky. For the 
UHECR, we consider 72 TA UHECR events with energies $E>57$ EeV using Surface Detector data collected between May 2008 and May 2013 
\cite{Abbasi14}. We have chosen the energy cut at $57$ EeV, since the backtracking method, in which our analysis relies on, is not able to 
deal with the large GMF deflections produced for energies lower than this limit. For the neutrino sample, we have used 12877 upward going 
muon neutrino candidate events recorded by IceCube in the 40-string configuration between April 2008 and May 2009 with a livetime of 375.5 
days \cite{Abbasi11}. While we are aware that there are more recent neutrino samples (e.g. \cite{Aartsen14a}), the arrival direction of 
individual events are not available, providing a distribution in zenith angle bins instead. 

For the 
purposes of this study we also work in Equatorial coordinates $(\alpha,\delta)$, where $0^\circ\leq \alpha < 360^\circ$ is the right ascension 
and $-90^\circ\leq \delta \leq 90^\circ$ is the declination. Given the geographical location of the IceCube neutrino observatory, the 
declination can be easily related to the zenith angle $\theta$ via the relation $\delta = 90 - \theta$.
For the galactic magnetic field scenario, we study four cases: no field, only with the regular JF12 field, the full JF12 field (regular plus
random components) and the standard turbulent field. 
 
In order to give a complete picture of our correlation analysis it is important to comment on the 
differences between the study in \cite{Martinez12} and ours. First, in contrast with the magnetic deflections used in \cite{Martinez12},
which are isotropic and independent of CR energy $E$, we are obtaining deflections as a function of $E$ and 
the arrival direction. The latter is an important difference since the deflections obtained via the backtracking method 
depend on the magnetic field traversed by the cosmic ray. 
Second, we are including the deflections from the regular field components that we use for shifting the cosmic ray arrival
positions, prior to using them for our correlation analysis. Third, we are considering in our study
a variety of GMF models. 

\subsection{SIGNAL AND BACKGROUND}
The expected number of signal neutrinos $\tilde{\mu}_s$ is given by:
\begin{equation}
\tilde{\mu}_s = T\int d\Omega \int dE_\nu A_{eff}(E_\nu,\textrm{cos}\theta) \Phi_\nu(E_\nu)
\end{equation}
where $T$ is the livetime, $A_{eff}$ is the IceCube muon-neutrino effective area in the 40-string configuration, $E_\nu$ is the 
neutrino energy, $\theta$ is the zenith angle, $\Phi$ is the muon-neutrino source flux and $d\Omega$ is the differential solid angle element. Effective areas are averaged 
over $30^\circ$ zenith bins as presented in \cite{Abbasi11}. Our hypotheses for the neutrino flux are $\Phi_\nu\propto E_\nu^{-2}$ 
and $\Phi_\nu\propto E_\nu^{-2.3}$, based on recent IceCube results \cite{Aartsen14b}. For a better understanding of 
our results it is important to remark that the $\tilde{\mu}_s$ is implicitly dependent on declinations since they are dependent on the 
zenith angle. 

To obtain the background we generate $10^6$ pseudo-experiments by scrambling the right ascension in the neutrino data. Centered
at each cosmic ray, we define a region of points within an angular distance $\Psi$ that we call angular bins of size $\Psi$. 
As mentioned before, we include deflections from the regular field components into the cosmic ray positions by shifting them accordingly. 
For each $\Psi$, the expected count $\mu_b(\Psi)$ of neutrinos and its standard deviation $\sigma_b$ are determined assuming a Gaussian 
distribution.

The value of $\Psi$ yielding the lowest 90\% Feldman-Cousins confidence level (CL) mean upper limit $\langle\mu_{90}\rangle$ 
\cite{Feldman98} is calculated, assuming background only. The value of $\langle\mu_{90}\rangle$ depends on $\mu_b$ only. We then calculate 
the expected signal $\tilde{\mu}_s$ based on Eq. (6), but adding the effects of angular resolution of the experiments by spreading the 
neutrino source according to a Gaussian distribution with standard deviation $2^\circ$. We also include an additional spread when dealing with 
the standard field and the JF12 random field components. We denote by ${\mu}_s$ the expected signal which embodies both effects (angular 
resolution and random magnetic field components). Because our magnetic field models are fixed before applying the analysis, there are no 
associated trial factors in the calculations. 

We find the angular bin size $\Psi_{MRF}$ which minimises the model rejection factor MRF=$\langle\mu_{90}\rangle/\mu_s$. For this 
$\Psi_{MRF}$, we determine the $5\sigma$ 90\% CL discovery potential $\mu_s^{5\sigma}$, calculated from the requirement 
$P(n\geq \mu_b+5 \sigma_b|\mu = \mu_b + \mu_s^{5\sigma}) = 0.5$, where $n$ is number of events. We then find the normalisation constant $C=E_\nu^\alpha\Phi_\nu$, which gives the 
strength of the source, from the equation $\mu_s^{5\sigma}= \mu_s(C)$ and solving for $C$. Note that while $\mu_s^{5\sigma}$ is the expected
neutrino count of signal neutrinos summed over all sources, the value of $C$ is the strength per source. 

\section{RESULTS}

\begin{figure}[h]
\centering
\includegraphics[width=0.5\textwidth]{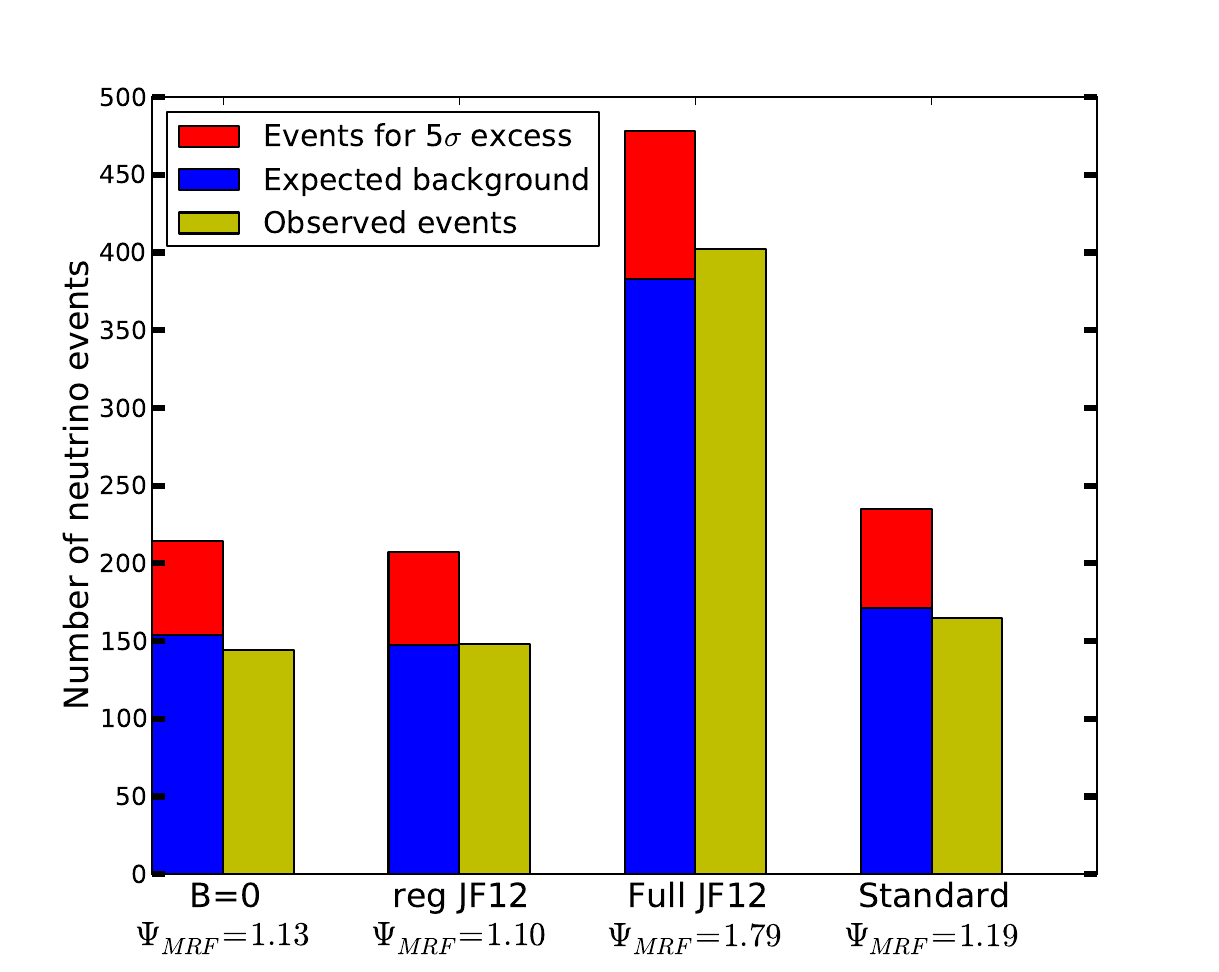}
\caption{Distribution of expected background, observed events and discovery potential for different magnetic field hypothesis.}
\label{Fig4}
\end{figure}

We find that for the four GMF model scenarios: no field (B=0), only with the regular JF12 field (reg. JF12), the full JF12 field consisting of 
the regular plus random components (full JF12) and the standard turbulent field, there is no 5$\sigma$ discovery after unblinding the data, 
being all the observed neutrino events consistent with the background. This result is expected since the chosen neutrino sample is consistent 
with atmospheric neutrinos and the compatibility between data and background is clearly shown in Figure \ref{Fig4}.  

In this way, introducing the full JF12 field does not improve the intensity of a correlation
signal between UHECR and neutrinos in comparison to other field assumptions. 
However, and in spite of this result, it is interesting to foresee the impact of choosing a particular GMF model in 
the values required to satisfy a positive correlation analysis. For that reason we consider worthwhile to 
extend our study in order to include the values of $\mu_s^{5\sigma}$ and the $90\%$ CL upper limits of 
the normalisation constant, which we obtain from the relation $\mu_{90}(\mu_b,n_{obs})=
\mu_s(C)$, where $\mu_{90}$ is calculated according to the Feldman-Cousins prescription.

In Tables 1 and 2 we show, for each of the GMF hypotheses, the values of $\Psi_{MRF}$, $\mu_s^{5\sigma}$ and the corresponding flux 
normalization constants $E_\nu^\alpha\Phi_\nu^{5\sigma}$ and $E_\nu^\alpha\Phi_\nu^{90}$ of $\mu_s^{5\sigma}$ and $90\%$ CL upper limits 
respectively. Tables 1 and 2 were obtained by using the 72 TA events and $\Phi_\nu\propto E_\nu^{-2}$/$\Phi_\nu\propto E_\nu^{-2.3}$ signal 
respectively. 

According to each magnetic field scenario, the behaviour of  $\Psi_{MRF}$ and $\mu_s^{5\sigma}$ are similar in the sense that, as $\Psi_{MRF}$ 
increases, so does $\mu_s^{5\sigma}$. This can be explained by the fact that $\mu_b$ and $\sigma_b$ increase with increasing $\Psi_{MRF}$, 
requiring a larger $\mu_s^{5\sigma}$ to get the desired $5\sigma$ excess above the expected background. The reason why we get a slightly 
higher $\mu_s^{5\sigma}$ for $B=0$ than the reg. JF12 case is that the change in declinations, after performing the magnetic field deflections
for the reg. JF12 component, turns out in a variation of $A_{eff}$ and thus in $\mu_s^{5\sigma}$. The normalisation constant 
$E_\nu^2\Phi_\nu^{5\sigma}$ is similar in both the $B=0$ and reg. JF12 cases because of the negligible change in $\Psi_{MRF}$. When we compare
the full JF12 against the standard scenario, we see that $\mu_s^{5\sigma}$ is stronger in the full JF12 but $E_\nu^2\Phi_\nu^{5\sigma}$ is weaker. This odd behaviour comes from the different treatment of the magnetic deflections in these models. As mentioned before, the standard
scenario assigns an energy independent deflection, with $\delta_\textrm{r.m.s.} = 3.0^\circ$, which tends to overestimate the random
deflections in comparison with the full JF12 (see Figure \ref{Fig3}). As a consequence, for the standard case, fewer neutrino events are
enclosed within an angular bin size $\Psi$ compared to the full JF12, thus the normalisation constant $E_\nu^2\Phi_\nu^{5\sigma}$ needs to be
higher in order to reach a given value of $\mu_s^{5\sigma}$. A similar relation is seen in Table 2.

It is also relevant to mention the quantitative differences between the full JF12 and standard field models. In Tables 1 and 2, $E_\nu^2\Phi_
\nu^{5\sigma}$ is smaller in the full JF12 by 15\%  and 17\% respectively, indicating that the strength of the point source required
for a $5\sigma$ discovery is somewhat overestimated in the case of the standard field. 

Regarding the dependence of the power law index $\alpha$ on our results, we observe that the relative differences between the normalization
constants for different field models are independent of $\alpha$. For the normalization constant itself we get higher values when 
$\alpha=2.3$. This can be explained by the steeply falling flux in this scheme, supressing contributions from the highest energy neutrinos at
the tail, which have the largest effective area and are the most relevant for $\mu_s$. The values of $\mu_s^{5\sigma}$ remain almost intact
after a change in $\alpha$, since the effect of this change is subdominant in our procedure. The results presented in Tables 1 and 2 are also
shown in Figure \ref{Fig4}, which is valid because of the negligible differences in $\Psi_{MRF}$ between both tables. The differences in the
upper limits is justified by the deviations of the observed numbered of events with respect to its expected background. If we observe a large
excess above the background, a larger flux per source consistent with the observed data are allowed. For this reason, in the full JF12
scenario, we have an upper limit $\sim 30\%$ higher than the other field models. 

Finally, in Table 3 we explore the effects on $\mu_s^{5\sigma}$ due to variations of the JF12 field model parameters and compare them to the results of 
our parameterization, assuming an $E_\nu^{-2}$ signal flux and using the 72 TA event data sample. We chose the parameters $B_0$ and $\beta$, 
where 
$B_0$ represents the magnetic field strength of the random halo component and $\beta$ modulates the strength of the random striated component 
with respect to the regular component via the relation $B_{stri}^2 = \beta B_{reg}^2$. There is an excellent agreement, of about 1\%, between
our parameterization and the JF12 numerical simulations, taking its best fit parameters. We achieve differences of at most $5\%$ after varying 
both parameters by one standard deviation.

\section{SUMMARY AND CONCLUSIONS} 
We have made a correlation analysis between the TA UHECR events and the IceCube-40 muon neutrino candidates under different GMF model 
assumptions. In particular, we have introduced the JF12 model, which expresses our best knowledge of the GMF. We have not found any 
correlations between the samples, regardless of the magnetic field hypothesis and the signal flux power law index, being the observed number 
of neutrino events consistent with the expected background. The JF12 full field model leads to a $90\%$ CL upper limit that is $\sim$ 30\% 
higher compared to the other models.	

We note that for the full JF12 field the angular bin size $\Psi_{MRF}$ is about 56\% higher than the other models, which in turn requires a 
higher number of signal events $\mu_s^{5\sigma}$ to claim a $5\sigma$ discovery. In contrast to other assumptions of the 
magnetic deflection, the JF12 field model incorporates an energy and arrival direction dependence that improves the accuracy of the analysis. 
In doing so, these deflections are not overestimated and leads to a signal strength per source that is weaker by $\sim 20\%$ for a 
$5\sigma$ discovery. The latter observation gives us the hope that weaker sources can yield positive 
correlations given enough time. The relative differences between our results should stay the same for 
future measurements. This analysis hints towards positive correlations due to relaxed flux limits per source
with further improvement of the knowledge of the GMF.

\section*{ACKNOWLEDGEMENTS}
The authors gratefully acknowledge DGI-PUCP for financial support under Grant No. 2014-0064, as well as CONCYTEC for graduate fellowship
under Grant No. 012-2013-FONDECYT. We also thank Esteban Roulet and Maria Teresa Dova for the useful discussions.

\begin{table*}
\begin{center}
\caption{$5\sigma$ 90\% CL discovery potential for the 72 TA events ($E_\nu^{-2}$ signal flux). $E_\nu^{2}\Phi_\nu^{5\sigma}$ and $E_\nu^2
\Phi_\nu^{90}$ are given in $\textrm{GeV cm}^{-2}\textrm{s}^{-1}\textrm{sr}^{-1}$.}
\begin{tabular}{ccccc}
\hline\hline
Field Model & $\Psi_{MRF}$ (deg) & $\mu_s^{5\sigma}$  & $E_\nu^2\Phi_\nu^{5\sigma}$ per source & $E_\nu^2\Phi_\nu^{90}$ per source\\
\hline
$B=0$ & 1.13 &  60.5 & $5.05\times 10^{-9}$ & $1.02\times 10^{-9}$\\
Reg. JF12 & 1.10 & 59.7 & $5.04\times 10^{-9}$ & $1.54\times 10^{-9}$\\
Full JF12 & 1.79 & 94.9 & $6.17\times 10^{-9}$ & $2.49\times 10^{-9}$\\
Standard & 1.19 & 63.6 & $7.27\times 10^{-9}$ & $1.93\times 10^{-9}$\\
\hline
\end{tabular}
\label{Table1}
\end{center}
\end{table*}

\begin{table*}
\begin{center}
\caption{Same as Table 1 for an $E_\nu^{-2.3}$ signal flux}
\begin{tabular}{ccccc}
\hline\hline
Field Model & $\Psi_{MRF}$ (deg) & $\mu_s^{5\sigma}$  & $E_\nu^2\Phi_\nu^{5\sigma}$ per source & $E_\nu^2\Phi_\nu^{90}$ per source\\
\hline
$B=0$ & 1.14 &  61.2 & $6.85\times 10^{-8}$ & $1.44\times 10^{-8}$\\
Reg. JF12 & 1.10 & 59.7 & $6.89\times 10^{-8}$ & $2.10\times 10^{-8}$\\
Full JF12 & 1.81 & 95.0 & $8.34\times 10^{-8}$ & $3.65\times 10^{-8}$\\
Standard & 1.19 & 63.6 & $1.00\times 10^{-7}$ & $2.64\times 10^{-8}$\\
\hline
\end{tabular}
\end{center}
\end{table*}

\begin{table*}
\begin{center}
\caption{$5\sigma$ 90\% CL discovery potential for the 72 TA events ($E_\nu^{-2}$ signal flux) under the Full JF12 Magnetic Field with varying
field parameters}
\begin{tabular}{ccccc}
\hline\hline
Field Model & $\Psi_{MRF}$ (deg)& $\mu_s^{5\sigma}$  & $E_\nu^2\Phi_\nu^{5\sigma}$ per source\\
\hline
Our work & 1.79 & 94.9 & $6.17\times 10^{-9}$\\
Best fit & 1.81 & 96.6 & $6.22\times 10^{-9}$\\
$\beta + \sigma$ & 1.80 & 95.3 & $6.22\times 10^{-9}$\\
$\beta - \sigma$ & 1.70 & 91.0 & $6.21\times 10^{-9}$\\
$B_0 + \sigma$ & 1.83 & 96.9 & $6.36\times 10^{-9}$\\
$B_0 - \sigma$ & 1.90 & 100.1 & $6.06\times 10^{-9}$\\
$\beta + \sigma, B_0 + \sigma$ & 1.99 & 104.7 & $6.37\times 10^{-9}$\\
$\beta - \sigma, B_0 - \sigma$ & 1.70 & 90.9 & $6.08\times 10^{-9}$\\
\hline
\end{tabular}
\end{center}
\label{Table4}
\end{table*}



\begin{thebibliography}{20}
\bibitem{Greisen66}
K. Greisen, Phys. Rev. Lett. \textbf{16}, 748 (1966).
\bibitem{Zatsepin66}
G.T. Zatsepin and V.A. Kuz'min, JETP Lett. \textbf{4}, 78 (1966).
\bibitem{Martinez12}
S. Adri\'an-Mart\'inez \textit{et al.} (ANTARES Collaboration), Astrophys. J. \textbf{774}, 19 (2012).
\bibitem{Fang14}
K. Fang, T. Fujii, T. Linden and A. Olinto, Astrophys. J (to be published). [accepted for publication]
\bibitem{Keivani15}
A. Keivani, G. Farrar and M. Sutherland, Astropart. Phys. \textbf{61}, 47 (2015).
\bibitem{Jansson12a}
R. Jansson and G.R. Farrar, Astrophys. J. \textbf{757}, 14 (2012).
\bibitem{Jansson12b}
R. Jansson and G.R. Farrar, Astrophys. J. \textbf{761}, L11 (2012).
\bibitem{Abbasi14}
R.U. Abbasi \textit{et al.} (Telescope Array Collaboration) Astrophys. J. \textbf{790}, L21 (2014).
\bibitem{Abbasi11}
R. Abbasi \textit{et al.} (IceCube Collaboration), Phys. Rev. D \textbf{84}, 082001 (2011).
\bibitem{Dolag04}
K. Dolag, D. Grasso, V. Springel and I. Tkachev, JETP Lett. \textbf{79}, 583 (2004).
\bibitem[Kotera \& Lemoine(2008)]{Kotera08}
K. Kotera and M. Lemoine, Phys. Rev. D \textbf{77}, 123003 (2008).
\bibitem{Lauer11}
R. Lauer \textit{et al.} (IceCube Collaboration), Astrophys. Space Sci. Trans. \textbf{7}, 201 (2011).
\bibitem{Aab14}
A. Aab \textit{et al.} (Pierre Auger Collaboration), Eur. Phys. J. C \textbf{75}, 269 (2015).
\bibitem{Gorski05}
K.M. G\'orski \textit{et al.}, Astrophys. J. \textbf{622} 758 (2005).
\bibitem{Batista13}
R. Alves Batista \textit{et al.}, in \textit{Proceedings of the 33rd International Cosmic Ray Conference, Rio de Janeiro, Brasil, 2013}.
\bibitem{Abreu10}
P. Abreu \textit{et al.} (Pierre Auger Collaboration), Astropart. Phys. \textbf{34}, 314 (2010).
\bibitem{Aartsen14a}
M.G. Aartsen \textit{et al.} (IceCube Collaboration), Phys. Rev. D, \textbf{89}, 062007 (2014).
\bibitem{Aartsen14b}
M.G. Aartsen \textit{et al.} (IceCube Collaboration), Phys. Rev. Lett. \textbf{113} 101101 (2014).
\bibitem{Feldman98}
G.J. Feldman and R.D. Cousins, Phys. Rev. D, \textbf{57}, 3873 (1998).
\end{thebibliography}


\end{document}